\documentclass[a4paper,twoside]{article}
\newcommand{\affil}[1]{$^{\rm #1}$}
\usepackage[authoryear]{natbib}
\usepackage{graphicx}
\date{} 
\title{\large\bf\flushleft The Globular Cluster System of NGC 5128:
  Ages, Metallicities, Kinematics, and Structural Parameters}
\author{\parbox{\textwidth}{\flushleft
\vspace{-0.5cm}
{\it K.A. Woodley\affil{A,B}, M. G{\'o}mez\affil{C}}\\
\vspace{0.4cm}
{\small \affil{A}\,Department of Physics \& Astronomy, McMaster
University, Hamilton ON  L8S 4M1, Canada}\\
{\small \affil{B}\,Current Address:  Department of Physics \& Astronomy, University of
  British Columbia, Vancouver BC V6T 1Z1, Canada.  Email: kwoodley@phas.ubc.ca}\\
{\small \affil{C}\,Departamento de Ciencias Fisicas, Facultad de Ingenieria, Universidad Andres Bello, Chile}\\
}}
\begin{document}
\twocolumn[
\begin{changemargin}{.8cm}{.5cm}
\begin{minipage}{.9\textwidth}
\vspace{-1cm}
\maketitle
%
%
\small{\bf Abstract:} 
We review our recent studies of the globular cluster system of NGC
5128.  First, we have obtained low-resolution, high signal-to-noise
spectroscopy of 72 globular clusters using Gemini-S/GMOS to obtain the ages,
metallicities, and the level of alpha enrichment of the
metal-poor and metal-rich globular cluster subpopulations.  Second, we have explored the
rotational signature and velocity dispersion of the galaxy's halo
using over 560 globular clusters with radial velocity measurements.
We have also compared the dependence of these properties on galactocentric distance
and globular cluster age and metallicity.  Using globular clusters as
tracer objects, we have analyzed the
mass, and M/L ratio of NGC 5128. Last, we have measured the structural
parameters, such as half-light radii, of over 570 globular
clusters from a superb 1.2 square degree Magellan/IMACS image.  
We will present the findings of these studies and
discuss the connection to the formation and evolution of NGC 5128.


\medskip
\medskip
\end{minipage}
\end{changemargin}
]
\small

\section{Introduction}

Globular clusters (GCs) survive for a Hubble time and
undergo dissipationless dynamical evolution.  They have been shown to
form during major episodes of star formation    \citep[][among
others]{holtzman92,schweizer93,whitmore93,whitmore95,zepf95,schweizer96,miller97,carlson98,schweizer98,whitmore99,zepf99,chien07,goudfrooij07,trancho07}.
This enables the use of GCs as probes of the formation history in
their host galaxy which can tell us important information on the
galaxy's assembly history and the physical conditions in which the GCs
were formed.  We can use their ages and metallicities to reconstruct
the star formation history of their galaxy \citep{west04} as well as
use the GCs as kinematic and dynamic tracers \citep{bridges06}.  

There are  many advantages in using  GCs as probes  of star formation
and galaxy assembly.  GCs are coeval and form with a single age and a single metallicity
to first approximation.  We also find  large GC  systems in
early-type galaxies, providing a large basis for study. GC systems
have been shown to exhibit two colour modes, both blue and red \citep{peng06},
indicating there could have been at least two episodes of star
formation within these galaxies.  Since the large majority of the
colour change of the  GC light that we see
happens within the first few Gyr \citep{worthey94}, any subsequent colour difference is
due to a difference in metallicity, so we refer to blue as metal-poor
and red as metal-rich GCs. 
The advancements of multi-object spectrographs have allowed us to
obtain numerous GC spectra in a homogeneous manner, providing large
sample sizes.

\subsection{The Globular Cluster System of NGC 5128}
NGC 5128 is the nearest available giant elliptical for study at $3.8
\pm0.1$ Mpc (G.L.H Harris, 2009, private communication).  Its close
proximity provides the opportunity to perform a detailed study of its
GC system at a level of detail that is not possible in other giant
ellipticals at present. We will use the GCs as representations of the
bulk stellar population to probe its formation history.

\begin{figure}[h]
\begin{center}
\includegraphics[scale=0.35, angle=0]{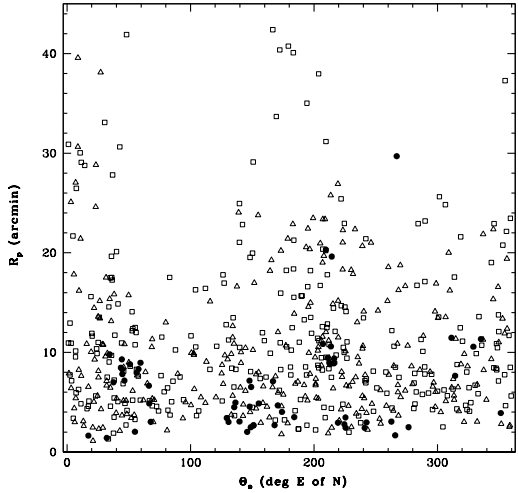}
\caption{The galactocentric radial distribution of GCs in NGC 5128 as
  function of azimuthal position, measured in degrees E of N. The {\it squares} and the {\it triangles} are
  the metal-poor and metal-rich GCs with a measured velocity, respectively, and the {\it solid circles}
are GCs that either have measured velocities but no colour information,
or GCs with no measured radial velocities.}\label{fig:Theta_R}
\end{center}
\end{figure}

In  NGC 5128, there are  an estimated  1500 GCs within 25 arcmin \citep{harris06}.  We
currently  know of  70 GCs  within NGC 5128 that have been identified
as resolved GCs from {\it Hubble Space Telescope} images
\citep{harris06}.  There are 564 GCs also confirmed by radial
velocity measurement
\citep{vandenbergh81,hesser84,hesser86,harris92,peng04b,woodley05,rejkuba07,beasley08},
including 190 new GCs our group has recently identified \citep{woodley09a,woodley09b}. 
The sample now includes 605 GCs which is one of  the  largest  samples  of GCs  in  the
literature. Of the total GC system with known photometry, we have
identified 268 metal-poor and  271 metal-rich
GCs, with a clear bimodal distribution in colour.  
Figure~\ref{fig:Theta_R} shows the projected spatial
distribution of the GC system which extends out to 45 arcmin in galactocentric
radius, distributed mainly along the isophotal major
axis of the  galaxy and heavily concentrated within 15 arcmin.  Our
sample is thus spatially biased, caused primarily by the chosen field
locations of previous studies.

\section{The Ages, Metallicities, and Alpha-Enhancement of the
  Globular Clusters}
\label{sec:ages}

\subsection{Index Measurement}

To obtain ages and metallicities of the GCs in NGC 5128, we need
to obtain high signal-to-noise (S/N) spectroscopy to measure the strength of particular
absorption features. 
In NGC 5128, we  have obtained  integrated  light
spectra of GCs with Gemini-S/GMOS.  Our field  placements are
centrally concentrated within 15 arcmin, but distributed azimuthally.
We obtained 
spectra for 72 GCs with S/N$>30/{\rm{\AA}}$ covering the wavelength regime of
$3800-5500\rm{\AA}$.   The measured indices include the Balmer lines,
as well as Mg, CN, and many Fe lines. 
The data reduction details are described in \cite{woodley09a}.

We measured the indices using GONZO \citep{puzia02} and
calibrated the data to the Lick index system \citep[described in][]{burstein84,worthey94,worthey97,
  trager98}, a standardized index system.  We had between 9--16 GCs in common with
\cite{beasley08} who had directly calibrated their data to the Lick index system via
standard stars.  With Lick indices in hand, we iterated between our
measured indices and the simple stellar population model grids of
\cite{tmb03} and \cite{tmk04} to obtain  ages,   metallicities,
and  alpha-to-iron abundance ratios,[$\alpha$/Fe], of the GCs.  We have performed the same analysis with
integrated light spectra of 40 Milky
Way GCs from \cite{schiavon05} and 1 Milky Way GC from \cite{puzia02}.
 We calibrated the Milky Way GC to the Lick system via 11 GCs
in common with \cite{puzia02}, who have calibrated their data to the Lick
index system with standard stars.  The uncertainties in our measured indices
are determined using GONZO by adding Poisson noise to each input GC
spectra via 100 Montecarlo simulations.  The uncertainty on each index
is the  $1\sigma$ standard deviation of the index measurement.  The
list of indices with uncertainties, as well as the various calibration plots are shown
in \cite{woodley09a}.

\subsection{Spectroscopic Results and Discussion}

\begin{figure}[h]
\begin{center}
\includegraphics[scale=0.35, angle=0]{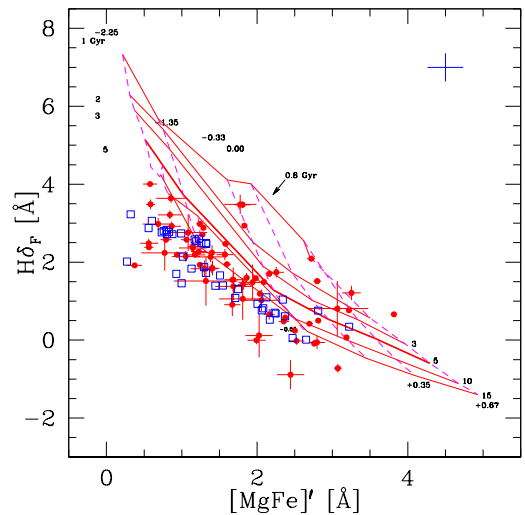}
\caption{The measured indices for 72 GCs in NGC 5128 ({\it red
    circles}) and 41 GCs in the Milky Way ({\it blue squares}) shown
  for H$\delta_F$ and [MgFe]$\prime$.  The systemic uncertainty is shown in the upper right
  ({\it blue cross}).  Overplotted are the simple stellar population
  models of \cite{tmb03,tmk04}. This figure is taken from \cite{woodley09a}. }\label{fig:hdeltaF}
\end{center}
\end{figure}

\begin{figure}[h]
\begin{center}
\includegraphics[scale=0.35, angle=0]{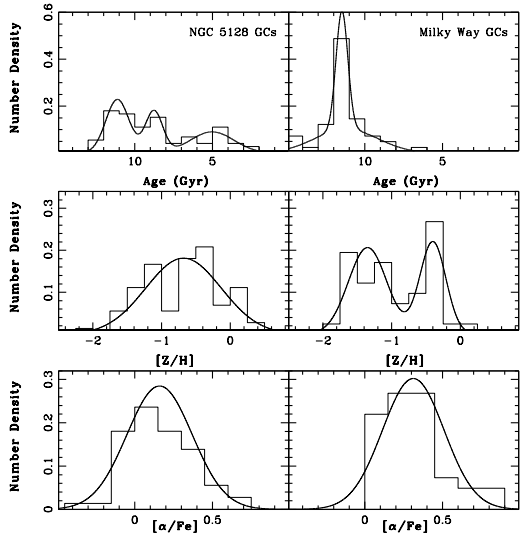}
\caption{The age ({\it top}), metallicity ({\it middle}), and
  [$\alpha$/Fe] ({\it bottom}) distributions for the GCs in NGC 5128
  ({\it left}) and the Milky Way ({\it right}).  The best fit
  Gaussians are plotted.  This figure has been modified from \cite{woodley09a}.}\label{fig:agemetafe}
\end{center}
\end{figure}

We show an example of one diagnostic plot in Figure~\ref{fig:hdeltaF}.
Here we have shown the GC indices for both NGC 5128 and the Milky Way plotted for the
higher order Balmer line H$\delta_F$ vs. [MgFe]$^\prime$ \footnote{[MgFe]$^\prime  = \sqrt{\rm{Mg}_b  \times\ (0.72  \times  \rm{Fe5270} +
0.28  \times \rm{Fe5335})}$  \citep{tmb03}}, sensitive to
age and metallicity, respectively.  Overplotted are the simple
stellar population models of \cite{tmb03} and \cite{tmk04} with an
[$\alpha$/Fe] of solar. To obtain age, metallicity, and [$\alpha$/Fe]
for the GCs, we use an iteration technique incorporating a number
of diagnostic plots, including the Balmer lines, Mg$_2$,  Mg$_b$,
Fe5270, and Fe5335 with  varying [$\alpha$/Fe] \citep[described in
full in][]{puzia02}. Our results are presented in Figure~\ref{fig:agemetafe}.  The
histrograms have been  fit with  Gaussian distributions  using the
statistical code  RMIX \footnote{ The  complete code, available  for a
variety of platforms, is publicly available from Peter MacDonald's Web
site at  http://www.math.mcmaster.ca/peter/mix/mix.html.}.  

We clearly see a difference in the age distribution functions between
the GCs in NGC 5128 and those in the Milky  Way. A Kolmogorov-Smirnov   statistical   comparison
test indicates these two distributions are different at greater than a
99\% confidence level.  There appears to be
multiple epochs for the formation history in NGC 5128.  An important test
of our technique is the ability to
reproduce the known old ages of the Milky Way GCs.  We have obtained a
mean age of  $11.3\pm0.1$ Gyr for the Milky Way GCs, in excellent
agreement with the known old ages of the Milky Way GCs.  We have
confidence therefore in our own results for the GCs in NGC 5128.
We find, in NGC 5128,  68$\%$
of  our GC  sample  have  old  ages greater than 8 Gyr,  14$\%$  have
intermediate ages between 5-8 Gyr, and 18$\%$ have young ages less
than 5 Gyr.  

The metallicity distribution, determined using spectroscopic indices
with the simple stellar population models, provides a distribution that
is not clearly bimodal, as seen in the Milky Way.  Our sample size is
quite small and is heavily biased towards metal-rich as well as
younger GCs because we are specifically targetting the central region of NGC 5128.
In this region, the metal-rich population is more strongly
concentrated than the metal-poor
\citep[see][]{woodley05,woodley07,beasley08}.  We are also more likely
to target younger objects, which are brighter than their older
counterparts, to achieve the require S/N.  We also have a bias
towards a larger range of
ages than perhaps the outer halo, due to accreted material sinking towards the
central regions.  Our proportion of young, metal-rich GCs is
therefore, likely inflated.   We find that  that 92$\%$ (23/25) of
metal-poor GCs and 56$\%$ (26/47) of metal-rich GCs that we have sampled in
NGC 5128 have ages $> 8$ Gyr, consistent with the old ages of the
Milky Way GCs.  We do however find all younger GCs are metal-rich,
with the exception of 2 metal-poor GC with an intermediate age. 

Lastly, we obtain the abundance of $\alpha$-to-Fe type elements.  This
ratio can indicate the timescale of GC formation.  Supernovae type II
events occur over a timescale of $\sim100$ Myr and enrich the
interstellar medium with primarily $\alpha$-type elements.  The onset of
supernova type Ia events, which enriches the interstellar medium with
a larger fraction of Fe-type elements after 1 Gyr, will reduce the value of 
[$\alpha$/Fe].  In NGC 5128, we find a spread in [$\alpha$/Fe] at
every given metallicity and a  wider  spread  among  the older GCs.
The mean [$\alpha$/Fe] for the GCs in NGC 5128 is $0.14\pm0.04$ dex which indicates a fast
formation timescale, likely forming in a rapid cloud collapse, as
opposed to a merger event, which would enrich the GCs to solar [$\alpha$/Fe].

\section{The Kinematics of the Globular Cluster System}

The bimodality of the GC system suggests there may be multiple epochs
of star formation.  This could also lead to different kinematic
signatures between the metal-rich and metal-poor GCs.   
We explore the kinematics signature of the 564 GCs with radial
velocity measurements by fitting the standard sine curve fit to the
data,
\begin{equation}
\label{eqn:kin}
v_p(\Theta) = v_{sys} + \Omega R  sin(\Theta - \Theta_o)
\end{equation}
described in \cite{cote01}.  Our input values are the measured radial  velocity,  $v_p$, and  the
angular  position of  each GC  measured on  the projected sky  in
degrees East of North, $\Theta$.  The radial  velocities  and  associated
uncertainties  used in  this study  are the  weighted averages  of all
previous measurements in the literature.  We extract the systemic
velocity,  $v_{sys}$,  the  rotation   amplitude, $\Omega  R$, and  the projected  rotation axis,
$\Theta_o$, also measured  in degrees East of North in  the projected
sky, with a weighted least squares non-linear fit.

We bin the data in two ways.  The first binning method is by radial position
in the galaxy so the bins are independent and plotted at
the mean radial value of all GCs in that bin.  The second binning method is an
exponentially weighted binning technique described in  \cite{bergond06}.

\subsection{Kinematic Results}

While we present only selected results here, the full results can be
found in \cite{woodley09b}.  Table~\ref{tab:kin} lists the $v_{sys}$,
$\Omega  R$, $\Theta_o$, and the velocity dispersion, $\sigma_{v_p}$,
for the metal-rich and metal-poor GC samples determined with Equation~\ref{eqn:kin}.

\begin{table}[h]
\begin{center}
\caption{Kinematics of the GC System}\label{tab:kin}
\begin{tabular}{lcc}
\hline  & Metal-Rich & Metal-Poor \\
\hline  \\
$v_{sys}$(km s$^{-1}$)    &  $527\pm10$&  $504\pm10$ \\
$\Omega R$(km s$^{-1}$)   &$ 41\pm15$  &$ 17\pm14$ \\
$\Theta_o$(deg E of N)   & $191\pm18$ &  $154\pm47$\\
$\sigma_{v_p}$(km s$^{-1}$)&$150\pm3 $ &$149\pm3 $\\
\hline
\end{tabular}
\medskip\\
\end{center}
\end{table}

We clearly see, as a whole, the GC system is
not strongly rotating and is thus supported by random motion.  The
metal-rich GCs appear to have mild rotation around an axis similar to
the isophotal major axis of the galaxy, located at 35 and 215 deg E of
N \citep{dufour79} in the inner 15 arcmin.  However, the metal-poor GCs with very mild rotation,
do not appear to rotate around either the major or minor isophotal
axis.  
The velocity dispersion is also determined for both the metal-rich and
metal-poor GC subpopulations, and are identical within
uncertainties.  This
is also seen in the GC system of the elliptical NGC  4636 \citep{schuberth06}.  

\begin{figure}[h]
\begin{center}
\includegraphics[scale=0.35, angle=0]{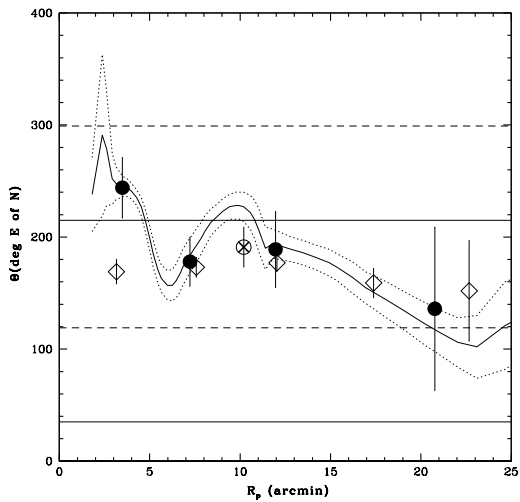}
\caption{The rotation axis as a function of galactocentric radius for
  the metal-rich GCs binned radially ({\it circles}) and exponentially
({\it curve}), and for the PNe ({\it triangles}).  The results for the
entire metal-rich population ({\it circle with x}) are also plotted. The isophotal
major ({\it solid lines}) and minor ({\it dashed lines}) axes are shown. }\label{fig:mr_kin_PASA}
\end{center}
\end{figure}

We examine the metal-rich GC population rotation axis more closely by
binning the data in independent galactocentric bins as well as by an
exponential fitting, shown in Figure~\ref{fig:mr_kin_PASA}.  For the
metal-rich GCs, the bin widths are 0-5, 5-10, 10-15, and 15-45 arcmin.
The rotation
axis of the metal-rich GCs does rotate around an axis near the isophotal major axis,
at least within the inner 10 arcmin of the galaxy. We also examine the kinematics for the age subsamples determined in
Section~\ref{sec:ages}, and are summarized in Table~\ref{tab:agekin}.
We note the
rotation axis of the metal-rich GCs with ages $< 5$ Gyr, do not rotate
around a similar axis as the bulk of old,  metal-rich GCs.  The
uncertainties are quite large  because
of the small number of young GCs in our sample (N=13), but it could
indicate the kinematics among the youngest GCs are different than
the bulk of the GC system.  A larger sample of GC ages are needed to
examine this more closely.  
\begin{table}[h]
\begin{center}
\caption{Kinematics of the Age Groups of GCs}\label{tab:agekin}
\begin{tabular}{lccc}
\hline  & $> 8$ Gyr & 5--8 Gyr & $< 5$ Gyr\\
\hline  \\
$v_{sys}$(km s$^{-1}$)      & 543$\pm$23&577$\pm$42  &562$\pm$51\\ 
$\Omega R$(km s$^{-1}$)    &27$\pm$36   &53$\pm$78  &58$\pm$92\\
$\Theta_o$(deg E of N)     & 107$\pm$65& 235$\pm$54 & 80$\pm$84\\
\hline
\end{tabular}
\medskip\\
\end{center}
\end{table}

\begin{figure}[h]
\begin{center}
\includegraphics[scale=0.35, angle=0]{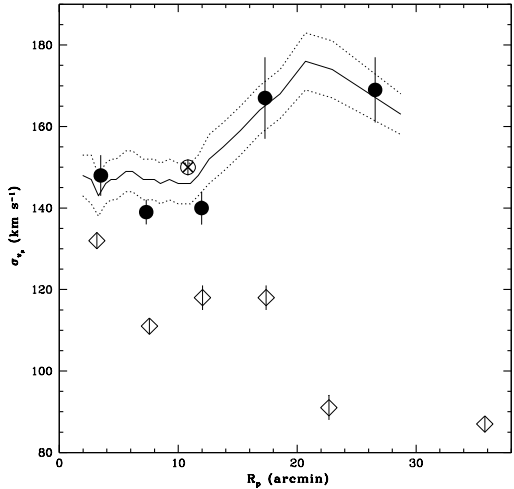}
\caption{The velocity dispersion as a function of galactocentric
  radius for the entire GC system binned radially ({\it circles}) and exponentially
({\it curve}), and for the PNe ({\it triangles}).  We also show the
results for the entire GC system ({\it circle with x}).}\label{fig:all_veldisp_pn}
\end{center}
\end{figure}

We also examine
the velocity dispersion for the entire GC system, as a function of
radius, shown in 
Figure~\ref{fig:all_veldisp_pn}.  In this plot, the radial bins are
0-5, 5-10, 10-15, 15-20, and 20-45 arcmin for the entire GC system.
The velocity dispersion appears flat
within the inner 15 arcmin, after which it shows an increase.
Although not shown in Figure~\ref{fig:all_veldisp_pn}, we note a
similar velocity dispersion profile for both the metal-rich GC and
metal-poor GC subpopulations separately \citep{woodley09b}.   This is
also the onset of the spatial biases within the GC system and needs to be
examined more closely with a larger, more azimuthally symmetric
sample of GCs.  With this caveat aside, an increasing velocity
dispersion in the outer regions could be interpreted as the galaxy
containing a large amount of dark matter or the GCs moving on
anisotropic orbits.  In NGC 5128 it is difficult to analyze the
orbital motion of the GCs without modelling and/or without an external
mass estimate extending to these outer regions.  Similar findings have
been noted in M87 \citep{cote01}, as well as hints for an increase in
the velocity dispersion of the GCs in the elliptical galaxies NGC 3379
\citep{pierce06} and M49 \citep{cote03}. 

\subsection{Comparison to the Planetary Nebulae and the Stellar Halo}

The planetary nebulae (PNe) are the most direct look at the field star
population in NGC 5128.  There is a sample of 780 PNe in NGC 5128
\citep{hui95,peng04a} extending out to $\sim80$ arcmin from the centre
of the galaxy, strongly concentrated along the well search isophotal
major axis.   We determine the rotation axis of the PNe and show its
profile distribution binned radially in
Figure~\ref{fig:mr_kin_PASA}. Similar to the metal-rich GCs, it
rotates around an axis near the isophotal major axis.  We
compare the PNe velocity dispersion to that of the GC system in
Figure~\ref{fig:all_veldisp_pn} and see a difference between the
two systems.  The bin sizes for the PNe data are 0-5,
5-10, 10-15, 15-20, 20-25, and 25-80 arcmin.  The PNe profile appears
to decrease, then flatten near
10--15 arcmin, and then decrease out to beyond 30 arcmin, where the GC
velocity dispersion appears to increase.  The decrease in velocity
dispersion of the planetary nebuale could indicate that the PNe are on
anisotropic orbits.  This behaviour has been seen in simulations of
merging galaxies \citep[for e.g.][]{dekel05} where the PNe get thrown
off their orbits from the disks of their progenitor galaxies into the
halo of the merger remnant.  We also note that the metal-rich GCs and
the PNe have very similar surface density profiles, with slopes of 
 $3.56\pm0.21$ and $3.47\pm0.12$ respectively, analyzed between 5--20
 arcmin, where 
we are most spatially secure with both sample populations.  
Further comparison with metallicity distribution of the stellar
halo population generated from {\it Hubble
  Space Telescope} data \citep[][and references therein]{rejkuba05},
shows a similar enrichment to the metal-rich GCs.  This is shown
graphically in \cite{woodley09a}.  \cite{rejkuba05} also estimate an
age for the stellar halo population of $\sim8^{+3}_{-3.5}$ Gyr, indicating the
stellar population, along with the results shown here for the
metal-rich GCs, are old.
These increasing similarities between the
metal-rich GCs and the stellar halo population could indicate that
they formed contemporaneously.  However, we cautiously note the
velocity dispersion profiles appear to differ between the metal-rich
GCs and the PNe.

\subsection{Mass and Mass-to-Light of NGC 5128}

GCs can also be used to trace the mass of its host galaxy.  Using our
sample of 564 GCs with radial velocity measurements, we use GCs as
tracer particles to determine a total mass of the galaxy.  The total mass is comprised of rotationally
supported mass added to the pressure supported mass.   We determine
the rotational component of the mass with the spherical Jeans
equation and the pressure supported mass with the Tracer Mass
Estimator \citep{evans03}.

Using the GCs that reside in the least spatially biased regime (between 5--20 arcmin),
we estimate a mass of NGC 5128 of $5.5\pm1.9 \times 10^{11}$ M$_{\odot}$ and a
mass-to-light ratio of 15.35 M$_{\odot}/$L$_{B\odot}$ \citep[assuming
B=7.84 for NGC 5128][]{karachentsev02} out
to 20 arcmin.  This mass estimate agrees quite well with other recent
work \citep{schiminovich94,peng04b,woodley07}.

\subsection{Kinematic Discussion}

For the large GC kinematic analysis presented in the literature, there
is a wide range of results.  
There are no clear patterns among the giant
ellipticals in the Virgo cluster of galaxies, M60
\citep{bridges06,hwang08}, M87 \citep{cote01}, M49 \citep{cote03}, and
NGC 4636 \citep{schuberth06}, as well as no clear patterns between the giant
elliptical galaxies near the dynamical centres of their cluster of
galaxies, M87 \citep{cote01} and NGC 1399 \citep{richtler04}, or the brightest ellipticals in their
host cluster of galaxies M49 \citep{cote03} and NGC 1399
\citep{richtler04}.  

\cite{bekki05} have numerically simulated the kinematic signatures of
preexisting metal-rich and metal-poor GCs from the merging of galaxies
like our Milky Way.  While their simulations were dissipationless, we are
interested in comparing their kinematic predictions to our findings,
since the bulk of the GC population in NGC 5128 is old.  In their
simulations, they found increasing rotational signatures with
increasing galactocentric radius from the merger, due to transfer to
angular momentum into the outer regions of the remnant.  We have
analyzed the rotation amplitude of NGC 5128 for both the metal-rich
and metal-poor GCs separately, and do not see a strong indication for
increasing rotation with radius.  The simulations of  \cite{bekki05} also
show the velocity dispersion for the GCs remained flat or
decrease with radius.  We see evidence for a flat dispersion out to
15 arcmin from the centre.  This comparison indicates that NGC
5128 may not be consistent with having formed as the result of
the merger of two fully formed disks galaxies, although this is not ruled
out.

An important implication from the study of
\cite{bekki05} suggests that the initial 
kinematic signatures of GCs in the progenitor galaxies undergo orbital
mixing.  It may not be possible to trace the {\it original}
kinematics of the GCs \citep{hwang08}, but only the GC kinematics from the most recent major
interaction. If this were the case, it would be very difficult to
use the current kinematics of the GCs to trace their orbital history
\citep{kisslerpatig98a}.
 
\section{Structural Parameters of the Globular Clusters}

The structural parameters of GCs can provide insight into how GCs form
as well as their environment of formation.  In particular, we are
interested in their half-light radii, $r_h$.  This parameter has been shown to
remain relatively constant throughout the lifetime of a GC
\citep{spitzer72,aarseth98}.   Interestingly, over the last decade,
many studies  have shown that the red GCs are smaller than the blue GCs
by 17--30\% in both spiral and elliptical galaxies
\citep{kundu98,kundu99,puzia99,larsen01a,larsen01b,kundu01,barmby02,jordan05,harris09}.
Almost all of these studies were performed with the small field of view of the {\it
  Hubble Space Telescope}, mainly
centred on the inner regions of their respective galaxies.
\cite{spitler06} examined the sizes of the GCs in the Sombrero galaxy
(NGC 4594), using the {\it Hubble Space Telescope}, but extending to a
few effective radii of the galaxy, R$_{eff}$.  They showed that the red GCs were smaller than
blue within the first 2  R$_{eff}$.  Beyond this distance their mean sizes did
not appear to differ.  

Our group has Magellan/IMACS images  taken in 0.45 arcsec seeing that
extend out to 40 arcmin in galactocentric radius.  With this stunning image,
 we can measure all GCs
with half-light diameters greater than 4 pc (0.22 arcsec).  In the
Milky Way, this
only misses 15\% of the GCs. These IMACS images give us the opportunity to study the trends
of the blue and red GCs homogeneously and extending to large radii.

We have measured the structural parameters with ISHAPE
\citep{larsen99,larsen01}  which  convolves  the  stellar  point 
spread  function with  an
analytical King  profile \citep{king62}  and compares the  result with
the  input candidate  image achieving  a best  match.   The structural
parameters,  measured from  the  models, are the  core
radius, $r_c$,  the tidal radius, $r_t$,  the concentration parameter,
$c=r_t/r_c$,  and  ellipticity.   The  half-light radii  can  also  be
obtained  from  the  transformation, $r_e/r_c  \simeq  0.547c^{0.486}$
which  is  good  to  $\pm2\%$  for $c>4$  \citep{larsen01},  which  is
satisfied for  GCs in NGC 5128 \citep[see][]{gomez07}.  We measured the
structural parameters for a total of 572 GCs in NGC 5128 and compare
the results for both red and blue GCs.  In this study, the
metallicity, [Fe/H], is a conversion from a (C-T$_1$) colour relation,
derived by \cite{harris02} calibrated through the
Milky Way GC data.  We use a  foreground reddening value of E(B - V) =
0.11  for NGC  5128 in this relation.  When there was no
available C-T$_1$ colour information for a GC, we converted their U-V
colour into a metallicity following the conversion of \cite{reed94},
calibrated with GCs in M31. 
We used a foreground reddening value of $E(U-V)= 0.2$ \citep{reed94}.  
Following \cite{harris04b}, \cite{woodley05}, and \cite{woodley07}, we divided our GC
sample into either metal-poor (when [Fe/H]$< -1.0$) or metal-rich (when 
[Fe/H]$ \geq -1.0$).  When neither C, T$_1$, U, or V information were
available, we assigned a GC as metal-rich if $(B-I)\geq 2.072$
and as metal-poor if $(B-I)< 2.072$ following \cite{peng04b}.

\begin{figure}[h]
\begin{center}
\includegraphics[scale=0.35, angle=0]{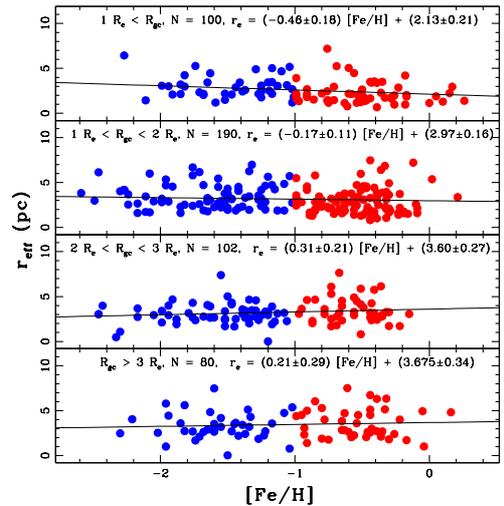}
\caption{The half-light radius for the GC system binned in effective
  radius of the galaxy, increasing in galactocentric distance
  from the top to bottom panels, as a function of metallicity [Fe/H].
This is plotted for both blue and red GCs (shown as {\it blue} and
{\it red circles}, respectively).  The data in each bin has a least
squares fit overplotted.}\label{fig:sp_fits}
\end{center}
\end{figure}

\begin{figure}[h]
\begin{center}
\includegraphics[scale=0.35, angle=0]{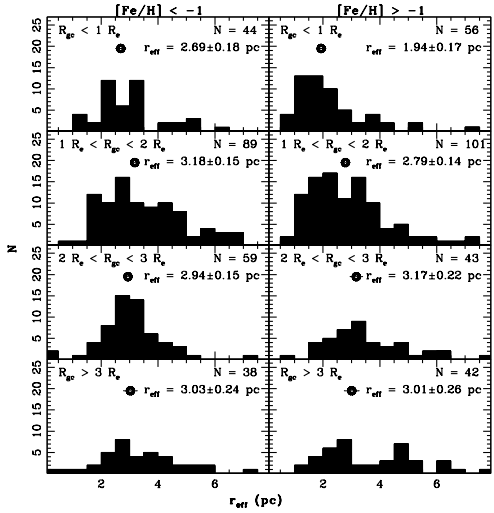}
\caption{The distributions of the half-light radii for the blue ({\it
    right}) and red ({\it left}) GCs, as a function of effective radii
in the galaxy.  The mean half-light radius in each bin is plotted with
uncertainty ({\it circles}).  }\label{fig:sp_histo}
\end{center}
\end{figure}

For the purpose of this study, we are interested in looking at the
sizes of the most representative GCs, and so we have removed the GCs that deviate from the
rest of the population in both ellipticity  and half-light
radii.  There are 472 GCs that satisfy the requirements of having both
ellipticity $<0.4$ and  $r_h<8$pc.  We display the measured $r_h$ in
Figure~\ref{fig:sp_fits} as a function of 
galactocentric radius.
The GCs are binned with effective galactic radius, R$_{eff}$, of
$\sim 5$ arcmin and fit each bin with a least squares fit.  We note
that within 1 R$_{eff}$, the blue GCs do appear to be larger than the
red, as has been seen in other GC systems, however, beyond 1--2
R$_{eff}$, this trend seems to disappear.
In Figure~\ref{fig:sp_histo}, we separate the blue GCs from the red GCs and plot
their half-light radii distributions as a function of the galaxy's
effective radius.  Along with the distributions, we show the mean
$r_h$ in each bin.  In the innermost bin, the mean half-light radius
for the blue GCs is $\sim30\%$ larger than for the red GCs.  Beyond this
distance the mean  $r_h$ in each bin is not distinguishable between
red and blue GCs.   This is consistent with the results we found in
\cite{gomez07} for a much smaller dataset of GCs in NGC 5128.
\subsection{Structural Parameter Discussion}

The size difference between the red and blue GCs is very interesting and it is
important to  understand why the red  GCs appear to be  smaller in the
inner regions  of galaxies.  There are a number of possible explanations
for this.

One explanation suggests it could be  caused by a projection  effect
\citep{larsen03}.  In the Milky  Way, it has been found that the GCs follow a size
–- distance relation, such that the size of a GC is smaller towards the
centre of  the galaxy \citep{vandenbergh91}.  If GC  systems in
other galaxies  follow this same  trend, then they suggest  that since
the MR  GCs are more  centrally concentrated in most extralactic systems,
they would  appear in the inner  regions to be smaller  than the blue GCs.
The  more  centrally  concentrated   red  GCs  would  lie  at  smaller
galactocentric distances  in projection, thus have  smaller radii.  At
larger distances, the projection would  have less effect and the trend
should dissappear.  However, recent work by \cite{harris09} who
studied a large combined sample of GCs from massive galaxies out to a few effective radii in each galaxy,  found the size difference between the red and blue GCs persisted at large distances.

The different sizes of GCs could be the combined effect of  mass
segregation  and   metallicity  dependent  stellar   lifetimes \citep{jordan05}. This would yield a smaller $r_h$ for metal-rich GCs, however, you would expect this to be the case at any
galactocentric radius.

The differences in sizes of blue and red GCs could be caused by
the different tidal
effects experienced during their formation \citep{harris09}.  In the  centre of
the galaxy, the potential well would be deeper than the outer regions,
and the GCs may have been forced to form smaller in order to survive.  This has been supported by recent work of
\cite{georgiev08,georgiev09} who measured sizes for the GCs in the Large
Magellanic Cloud and noticed they were larger than the normal GC population in the Milky Way.   
If the size difference in the inner regions fo the galaxy is real,
then this could suggest that metal-rich and metal-poor GCs in some
extragalactic systems, did not form under the same conditions. 

We require more homogeneous and large samples of GCs in a variety of
galaxies in order to better understand what processes in GC
formation or formation environment has on the formation sizes of GCs. 
Interestingly, the size differences of GCs may have a direct
relation to the environment in which they form, allowing the
possibility of GCs to probe formation history of their host galaxy in
yet another way.  We suggest that it is likely the combination of the
environment that the GCs have formed with a projection
effect that may be responsible for the size differences in the inner
regions of the GCs within NGC 5128.

\section{Conclusions}

We have studied the GC system in NGC 5128 and presented
their ages and metallicities, their kinematics, their structural
parameters, and we have also used them as tracers to estimate the mass 
of NGC 5128.

We find, using a sample of 72 GCs, that the majority of metal-rich and
metal-poor GCs are old with 68\% having ages $> 8$ Gyr.  We do find a
small fraction with young ages $< 5$ Gyr, all of which are metal-rich.
These results suggest there could have been a number of formation
epochs within NGC 5128, which resulted in the formation of a small
portion of its stellar population.  

Using over 560 GCs in NGC 5128, we performed a kinematic analysis, and
find the metal-rich GCs have a mild rotation of $ 41\pm15$ km s$^{-1}$ and rotate around
$191\pm18$ deg E of N, similar to the isophotal major axis of the galaxy.  The metal-rich
GCs have very mild rotation and do not appear to rotate around either
isophotal major or minor axis.  The velocity dispersion profile of the
GCs is
generally flat in the inner regions of the galaxy.  Compared to the
stellar halo population, we see a number of similarities to the
metal-rich GCs, including the rotation axis and surface density
profile of the PNe population, as well as approximate ages and metal
enrichment of the stellar halo \citep{rejkuba05}.  We also use the
radial velocity measurements of the GCs to estimate a mass of NGC 5128
to be
$5.5\pm1.9 \times 10^{11}$ M$_{\odot}$ and a
mass-to-light ratio of 15.35 M$_{\odot}/$L$_{B\odot}$ out to 20 arcmin.

We examine the structural parameters of a representative sample of
normal GCs in NGC 5128 using superb IMACS images.  We examine their
half-light radii, and find within 1--2 effective radii of the galaxy,
the blue GCs are $\sim30\%$ larger than the red GCs.  Beyond this
distance, however, the size difference seems to disappear.  We suggest
this may be the result of their environment of formation, combined with
the projection effect.

The old ages and kinematics of the GCs within NGC 5128 suggests that 
the majority of GCs, and
presumably stars, formed early on in a rapid, or perhaps multiple,
collapse(s).  This is supported by the old ages of the GCs and high
[$\alpha$/Fe] for both metal-rich and metal-poor GCs.  There is
evidence for a large spread in metal-rich ages, indicating either
minor merging or accretion of small neighbouring satellites in more
recent times.  However the fraction of young objects is heavily biased by
our selection of targets.  The youngest GCs that we find in our sample are
a few Gyrs old.   

\section*{Acknowledgments} K.A.W. would like to thank the science
organizing committee for the opportunity to present this work at the
Many Faces of Centaurus A Conference in Sydney as well as for
financial support. M.G. thanks Proyecto DI-36-09/R for financial support.



\begin{thebibliography}{}
\bibitem[Aarseth et al.(1998)]{aarseth98} Aarseth, S.~J. \& Heggie,
  D.~C. 1998, MNRAS, 297, 794
\bibitem[Barmby et al.(2002)]{barmby02} Barmby, P., Holland, S., \& Huchra, J. P. 2002, AJ, 123, 1937
\bibitem[Beasley et al.(2008)]{beasley08} Beasley, M.~A., Bridges, T.,
  Peng, E., Harris, W.~E., Harris, G.~L.~H., Forbes, D.~A., \& Mackie,
  G. 2008, MNRAS, 386, 1443
\bibitem[Bekki et al.(2005)]{bekki05} Bekki, K., Beasley, M.~A.,
  Brodie, J.~P., \& Forbes, D.~A. 2005, MNRAS, 363, 1211
\bibitem[Bergond et al.(2006)]{bergond06} Bergond, G., Zepf, S.~E.,
  Romanowsky, A.~J., Sharples, R.~M., \& Rhode, K.~L. 2006, A\&A, 448,
  155
\bibitem[Bridges et al.(2006)]{bridges06} Bridges, T., Gebhardt, K.,
  Sharples, R., Faifer, F.~R., Forte, J.~C., Beasley, M.~A., Zepf,
  S.~E., Forbes, D.~A., Hanes, D.~A., \& Pierce, M. 2006, MNRAS, 373,
  157
\bibitem[Burstein et al.(1984)]{burstein84} Burstein, D., Faber,
  S. M., Gaskell, C. M., \& Krumm, N. 1984, ApJ, 287, 586 
\bibitem[Carlson et al.(1998)]{carlson98} Carlson, M. N., Holtzman,
  J. A., Watson, A. M., Grillmair, C. J., Mould, J. R., Ballester,
  G. E., Burrows, C. T., Clarke, J. T., Crisp, D., Evans, R. W.,
  Gallagher, J. S. III, Griffiths, R. E., Hester, J. J., Hoessel,
  J. G., Scowen, P. A., Stapelfeldt, K. R., Trauger, J. T., \&
  Westphal, J. A. 1998, AJ, 115, 1778
\bibitem[Chien et al.(2007)]{chien07} Chien, L. -H., Barnes, J. E.,
  Kewley, L. J., \& Chambers, K. C., 2007, ApJ, 660, 105
\bibitem[C{\^o}t{\'e} et al.(2001)]{cote01} C{\^o}t{\'e}, P.,
  McLaughlin, D.~E., Hanes, D.~A., Bridges, T.~J., Geisler, D.,
  Merritt, D., Hesser, J.~E., Harris, G.~L.~H., \& Lee, M.~G. 2001,
  ApJ, 559, 828
\bibitem[C{\^o}t{\'e} et al.(2003)]{cote03} C{\^o}t{\'e}, P.,
  McLaughlin, D.~E., Cohen, J.~G., \& Blakeslee, J.~P. 2003, ApJ,591,850
\bibitem[Dekel et al.(2005)]{dekel05} Dekel, A., Stoehr, F., Mamon,
  G.~A., Cox, T.~J., Novak, G.~S., \& Primack, J.~R. 2005, Nature,
  437, 707
\bibitem[Dufour et al.(1979)]{dufour79} Dufour, R.~J., Harvel, C.~A.,
  Martins, D.~M., Schiffer, III, F.~H., Talent, D.~L., Wells, D.~C.,
  van den Bergh, S., \& Talbot, Jr., R.~J. 1979, AJ, 84, 284
\bibitem[Evans et al.(2003)]{evans03} Evans, N.~W., Wilkinson, M.~I.,
  Perrett, K.~M., \& Bridges, T.~J. 2003, ApJ, 583, 752 
\bibitem[Georgiev et al.(2008)]{georgiev08} Georgiev, I. Y., Goudfrooij, P., Puzia, T. H., \& Hilker, M. 2008, AJ, 135, 1858 
\bibitem[Georgiev et al.(2009)]{georgiev09}Georgiev, I. Y., Puzia, T. H., Hilker, M., \& Goudfrooij, P. 2009, MNRAS, 392, 879 
\bibitem[Gom{\'e}z \& Woodley(2007)]{gomez07} Gom{\'e}z, M. \&
  Woodley, K. A. 2007, ApJ, 670, L105
\bibitem[Goudfrooij et al.(2007)]{goudfrooij07} Groudfrooij, P.,
  Schweizer, F., Gilmore, D., \& Whitmore, B. C. 2007, AJ, 133, 2737
\bibitem[Harris et al.(1992)]{harris92} Harris, G.~L.~H., Geisler, D.,
  Harris, H.~C., \& Hesser, J.~E. 1992, AJ, 104, 613
\bibitem[Harris \& Harris(2002)]{harris02} Harris, W.~E. \& Harris,
  G.~L.~H. 2002, AJ, 123, 3108
\bibitem[Harris et al.(2004)]{harris04b} Harris, G.~L.~H., Harris,
  W.~E., \& Geisler, D. 2004, AJ, 128, 723
\bibitem[Harris et al.(2006)]{harris06} Harris, W. E., Harris,
  G. L. H., Barmby, P., McLaughlin, D. E., \& Forbes, D. A. 2006, AJ,
  132, 2187
\bibitem[Harris(2009)]{harris09} Harris, W. E. 2009, ApJ,  699, 254
\bibitem[Hesser et al.(1984)]{hesser84} Hesser, J.~E., Harris, H.~C.,
  van den Bergh, S., \& Harris, G.~L.~H. 1984, ApJ, 276, 491
\bibitem[Hesser et al.(1986)]{hesser86} Hesser, J.~E., Harris, H.~C.,
  \& Harris, G.~L.~H. 1986, ApJL, 303, 51
\bibitem[Holtzman et al.(1992)]{holtzman92} Holtzman, J. A., Faber,
  S. M., Shaya, E. J., Lauer, T. R., Groth, J., Hunter, P. A., Baum,
  W. A., Ewald, S. P., Hester, J. J., Light, R. M., Lynds, C. R.,
  O'Neil, E. J. Jr., \& Westphal, J. A. 1992, AJ, 103,691 
\bibitem[Hui et al.(1995)]{hui95} Hui, X., Ford, H.~C., Freeman,
  K.~C., \& Dopita, M.~A. 1995, ApJ, 449, 592
\bibitem[Hwang et al.(2008)]{hwang08} Hwang, H.~S., Lee, M.~G., Park,
  H.~S., Kim, S.~C., Park, J.-H., Sohn, Y.-J., Lee, S.-G., Rey, S.-C.,
  Lee, Y.-W., \& Kim, H.-I. 2008, ApJ, 674, 869
\bibitem[Jord{\'a}n et al.(2005)]{jordan05} Jord{\'a}n, A.,
  C{\^o}t{\'e}, P., Blakeslee, J. P., Ferrarese, L., McLaughlin,
D. E., Mei, S., Peng, E. W., Tonry, J. L., Merritt, D.,
Milosavljevi{\'c}, M., Sarazin, C. L., Sivakoff, G. R., \& West, M. J. 2005, ApJ, 634, 1002
\bibitem[Karachentsev et al.(2002)]{karachentsev02} Karachentsev,
  I.~D., Sharina, M.~E., Dolphin, A.~E., Grebel, E.~K., Geisler, D.,
  Guhathakurta, P., Hodge, P.~W., Karachentseva, V.~E., Sarajedini,
  A., \& Seitzer, P. 2002, A\&A, 385, 21
\bibitem[King(1962)]{king62} King, I. R. 1962, AJ, 67, 471
\bibitem[Kissler-Patig et al.(1998)]{kisslerpatig98a} Kissler-Patig,
  M., Brodie, J.~P., Schroder, L.~L., Forbes, D.~A., Grillmair, C.~J.,
  \& Huchra, J.~P. 1998, AJ, 115, 105
\bibitem[Kundu \& Whitmore(1998)]{kundu98} Kundu, A. \& Whitmore,
  B. C. 1998, AJ, 116, 2841
\bibitem[Kundu \& Whitmore(2001)]{kundu01} Kundu, A. \& Whitmore,
  B. C. 2001, AJ, 121, 2950 
\bibitem[Kundu et al.(1999)]{kundu99}  Kundu, A., Whitmore, B. C.,
  Sparks, W. B., Macchetto, R. D., Zepf, S. E., \& Ashman, K. M. 1999,
  ApJ, 513, 733
\bibitem[Larsen(1999)]{larsen99} Larsen, S. S. 1999, A\&ApS, 139, 393
\bibitem[Larsen(2001)]{larsen01} Larsen, S. S. 2001, AJ, 122, 1782
\bibitem[Larsen \& Brodie(2003)]{larsen03} Larsen, S. S., \& Brodie, J. P. 2003, ApJ, 593, 340
\bibitem[Larsen et al.(2001a)]{larsen01a} Larsen, S. S., Brodie, J. P., Huchra, J. P., Forbes, D. A., \& Grillmair, C. J. 2001a, AJ, 121, 2974
\bibitem[Larsen et al.(2001b)]{larsen01b} Larsen, S. S., Forbes, D. A., \& Brodie, J. P. 2001b, MNRAS, 327, 1116 
\bibitem[Miller et al.(1997)]{miller97} Miller, B. W., Whitmore,
  B. C., Schweizer, F., \& Fall, S. M. 1997, AJ, 114, 2381
\bibitem[Peng et al.(2004a)]{peng04a} Peng, E.~W., Ford, H.~C., \&
  Freeman, K.~C. 2004a, 602, 685
\bibitem[Peng et al.(2004b)]{peng04b} Peng, E.~W., Ford, H.~C., \&
  Freeman, K.~C. 2004b, ApJ, 602, 705
\bibitem[Peng et al.(2006)]{peng06} Peng, E.~W., Jord{\'a}n, A.,
  C{\^o}t{\'e}, P., Blakeslee, J.~P., Ferrarese, L., Mei, S., West,
  M.~J., Merritt, D., Milosavljevi{\'c}, M., \& Tonry, J.~L. 2006,
  ApJ, 639, 95
\bibitem[Pierce et al.(2006)]{pierce06} Pierce, M., Beasley, M.~A.,
  Forbes, D.~A., Bridges, T., Gebhardt, K., Faifer, F.~R., Forte,
  J.~C., Zepf, S.~E., Sharples, R., Hanes, D.~A., \& Proctor, R. 2006,
  MNRAS, 366, 1253
\bibitem[Puzia et al.(1999)]{puzia99} Puzia, T. H., Kissler-Patig, M., Brodie, J. P., \& Huchra, J. P. 1999, AJ, 118, 2734
\bibitem[Puzia et al.(2002)]{puzia02} Puzia, T. H., Zepf, S. E.,
  Kissler-Patig, M., Hilker, M., Minniti, D., \& Goudfrooij, P. 2002,
  A\&A, 391, 453
\bibitem[Reed et al.(1994)]{reed94} Reed, L.~G., Harris, G.~L.~H., \&
  Harris, W.~E. 1994, AJ, 107, 555
\bibitem[Rejkuba et al.(2005)]{rejkuba05} Rejkuba, M., Greggio, L.,
  Harris, W.~E., Harris, G.~L.~H., \& Peng, E.~W. 2005, 631, 262
\bibitem[Rejkuba et al.(2007)]{rejkuba07} Rejkuba, M., Dubath, P.,
  Minniti, D., \& Meylan, G. 2007, A\&A, 469, 147
\bibitem[Richtler et al.(2004)]{richtler04} Richtler, T., Dirsch, B.,
  Gebhardt, K., Geisler, D., Hilker, M., Alonso, M.~V., Forte, J.~C.,
  Grebel, E.~K., Infante, L., Larsen, S., Minniti, D., \& Rejkuba,
  M. 2004, AJ, 127, 2094
\bibitem[Schiminovish et al.(1994)]{schiminovich94} Schiminovich, D.,
  van Gorkom, J.~H., van der Hulst, J.~M., \& Kasow, S. 1994, ApJL,
  423, 101
\bibitem[Schweizer \& Seitzer(1993)]{schweizer93} Schweizer, F. \&
  Seitzer, P. 1993, ApJ, 417, L29
\bibitem[Schweizer et al.(1996)]{schweizer96} Schweizer, F., Miller,
  B. W., Whitmore, B. C., \& Fall, S. M. 1996, AJ, 112, 1839
\bibitem[Schiavon et al.(2005)]{schiavon05} Schiavon, R. P., Rose,
  J. A., Courteau, S., \& MacArthur, L. A. 2005, ApJS, 160, 163
\bibitem[Schuberth et al.(2006)]{schuberth06} Schuberth, Y., Richtler,
  T., Dirsch, B., Hilker, M., Larsen, S.~S., Kissler-Patig, M., \&
  Mebold, U. 2006, A\&A, 459, 391
\bibitem[Schweizer \& Seitzer(1998)]{schweizer98} Schweizer, F. \&
  Seitzer, P. 1998, AJ, 116, 2206
\bibitem[Spitler et al.(2006)]{spitler06} Spitler, L. R., Larsen,
  S. S., Strader, J., Brodie, J. P., Forbes, D. A., \& Beasley,
  M. A. 2006, AJ, 132, 1593
\bibitem[Spitzer \& Thuan(1972)]{spitzer72} Spitzer, L.~J., \& Thuan,
  T.~X. 1972, ApJ, 175, 31
\bibitem[Thomas et al.(2003)]{tmb03} Thomas, D., Maraston, C., \&
  Bender, R. 2003, MNRAS, 339, 897
\bibitem[Thomas et al.(2004)]{tmk04} Thomas, D., Maraston, C., \&
  Korn, A. 2004, MNRAS, 351, L19
\bibitem[Trager et al.(1998)]{trager98} Trager, S. C., Worthey, G.,
  Faber, S. M., Burstein, D., \& Gonzalez, J. J. 1998, ApJS, 116, 1
\bibitem[Trancho et al.(2007)]{trancho07} Trancho, G., Bastian, N.,
  Miller, B. W., \& Schweizer, F. 2007, ApJ, 664, 284
  M. J. 1978, MNRAS, 183, 341
\bibitem[van den Bergh et al.(1981)]{vandenbergh81} van den Bergh, S.,
    Hesser, J.~E., \& Harris, G.~L.~H. 1981, AJ, 86, 24
\bibitem[van den Bergh et al.(1991)]{vandenbergh91} van den Bergh, S., Morbey, C., \& Pazder, J. 1991, ApJ, 375, 594
\bibitem[West et al.(2004)]{west04} West, M.J., C{\'o}t{\^e}, P.,
  Marzke, R.O., Jord{\'a}, A. 2004, Nature, 427, 31
\bibitem[Whitmore et al.(1993)]{whitmore93} Whitmore, B. C.,
  Schweizer, F., Leitherer, C., Borne, K., Robert, C. 1993, AJ, 106, 1354
\bibitem[Whitmore \& Schweizer(1995)]{whitmore95} Whitmore, B. C. \&
  Schweizer, F. 1995, AJ, 109, 960
\bibitem[Whitmore et al.(1999)]{whitmore99} Whitmore, B. C., Zhang,
  Q., Leitherer, C., Fall, S. M., Schweizer, F., \& Miller,
  B. W. 1999, AJ, 118, 1551
\bibitem[Woodley et al.(2005)]{woodley05} Woodley, K.~A., Harris,
  W.~E., \& Harris, G.~L.~H. 2005, AJ, 129, 2654
\bibitem[Woodley et al.(2007)]{woodley07} Woodley, K.~A., Harris,
  W.~E., Beasley, M.~A., Peng, E.~W., Bridges, T.~J., Forbes, D.~A.,
  \& Harris, G.~L.~H. 2007, 134, 494
\bibitem[Woodley et al.(2009a)]{woodley09a} Woodley, K. A., Harris,
  W. E., Puzia, T. H., G{\'o}mez, M., Harris, G. L. H., \&  Geisler,
  D. 2009a, ApJ, accepted
\bibitem[Woodley et al.(2009b)]{woodley09b} Woodley, K. A., Harris,
  W. E., G{\'o}mez, M., Geisler, D. \& Harris, G. L. H., 2009b, AJ, submitted
\bibitem[Worthey \& Ottaviani(1997)]{worthey97} Worthey, G. \&
  Ottaviani, D. L. 1997, ApJS, 111, 377
\bibitem[Worthey et al.(1994)]{worthey94} Worthey, G., Faber, S. M.,
  Gonzalez, J. J., \& Burstein, D. 1994, ApJS, 94, 687
\bibitem[Zepf et al.(1995)]{zepf95} Zepf, S. E., Carter, D., Sharples,
  R. M., \& Ashman, K. 1995, ApJ, 445, L19
\bibitem[Zepf et al.(1999)]{zepf99} Zepf, S. E., Ashman, K. M.,
  English, J., Freeman, K. C., \& Sharples, R. M. 1999, AJ, 118, 752


\end{thebibliography}
\end{document}